# Risk Analysis in the Selection of Project Managers Based on ANP and FMEA

**Armin Asaadi**, Department of Industrial Management, Faculty of Management, Firoozkooh branch, Islamic Azad University, Firoozkooh, Iran. a19880213asaadi@gmail.com

**Armita Atrian,** Department of Management, Faculty of Administrative Sciences, Imam Reza International University, Mashhad, Iran. armita.atrian@imamreza.ac.ir

**Hesam Nik Hoseini,** Department of Industrial Management, Faculty of Management, Science and Research Branch, Islamic Azad University, Firoozkooh, Iran.

**Mohammad Mahdi Movahedi,** Department of Industrial Management, Faculty of Management, Firoozkooh branch, Islamic Azad University, Firoozkooh, Iran.

**Abstract**

Project managers play a crucial role in the success of projects. The selection of an appropriate project manager is a primary concern for senior managers in firms. Typically, this process involves candidate interviews and assessments of their abilities. There are various criteria for selecting a project manager, and the importance of each criterion depends on the project type, its conditions, and the risks associated with their absence in the chosen candidate. Often, senior managers in engineering companies lack awareness of the significance of these criteria and the potential risks linked to their absence. This research aims to identify these risks in selecting project managers for civil engineering projects, utilizing a combined ANP-FMEA approach. Through a comprehensive literature review, five risk categories have been identified: individual skills, power-related issues, knowledge and expertise, experience, and personality traits. Subsequently, these risks, along with their respective sub-criteria and internal relationships, were analysed using the combined ANP-FMEA technique. The results highlighted that the lack of political influence, absence of construction experience, and deficiency in project management expertise represent the most substantial risks in selecting a project manager. Moreover, upon comparison with the traditional FMEA approach, this study demonstrates the superior ability of the ANP-FMEA model in differentiating risks and pinpointing factors with elevated risk levels.

## 1. Introduction

In the world of project management, employees are the heart and soul of any successful endeavour. Their passion, dedication, and unique talents form the foundation upon which projects thrive. It's the creativity and expertise they bring to the table that breathes life into ideas and turn plans into reality. Collaboration among team members, each with their own experiences and skills, not only leads to innovative solutions but also creates a vibrant and dynamic work environment. The ability to communicate effectively and work harmoniously ensures that everyone is on the same page, working towards common goals. Beyond just tasks and deadlines, it's the enthusiasm and commitment of these individuals that infuse a sense of camaraderie and motivation into the project space (Daneshmandi et al., 2023). Their contributions go beyond the tangible—they create a supportive and encouraging atmosphere that fuels not only project efficiency but also a sense of belonging and shared achievement. In essence, employees are the unsung heroes, shaping projects with their humanity, passion, and collaboration, making each endeavour a true success story (Hessari & Nategh, 2022b). A project can simply be defined as a unique collection of activities with a specific end time. With this straightforward definition, it becomes evident that many activity collections in various fields, such as building a ship, modifying a car part, constructing a new building, and more, can be classified as projects. Traditionally, projects have been managed using a functional organizational structure, which is hierarchical and task oriented (Zhu et al., 2022). However, as projects become more complex, such as multinational and multi-company projects, projects involving collaboration from different sections of a company, and projects executed with the involvement of various specialties, companies are inclined towards matrix structures, project teams, and project management (Jayasudha & Vidivelli, 2016).

In recent years, the number of construction projects has significantly increased, involving multiple decision-making stages throughout the project. Decision-making is always intricate and challenging, especially when multiple criteria need to be considered. Decisions such as choosing appropriate implementation methods, selecting suitable materials, and workforce are all tasks that must be executed by the manager (Hessari et al., 2022a). However, perhaps the most crucial decision in this context is the selection of the project manager, as they play a significant role in subsequent decision-making processes (Ahmed & El-Sayegh, 2020). Since choosing a manager is always accompanied by risks due to the limited resources available and the intricate process of designing and constructing a project, selecting the most suitable project manager is of paramount importance (Agbejule & Lehtineva, 2022).

Different projects require project managers with varying skills and capabilities. Consequently, criteria for selecting a manager are established based on the project's specific needs (Hessari & Nategh, 2022c). These criteria often lack clear and definitive values, making decision-making in such contexts challenging and risky. Furthermore, the process of selecting human resources usually occurs informally and without a structured method, allowing room for errors (Ahmadi-Javid et al., 2020). These uncertain judgments are not easily expressible through precise numerical values. Therefore, to address the complexities of decision-making in such matters, employing new interdisciplinary approaches is essential. Since the success of projects, both in the design and construction phases, significantly relies on the capabilities of the project manager, methods for enhancing decision-making and selecting the most suitable project manager are crucial. Considering that the project manager holds a singular position of responsibility, it is their duty to establish a structure that caters to the project's needs, the organization, stakeholders, and project personnel (Cakmak & Tezel, 2019).

This article focuses on the risk associated with selecting a competent manager for construction projects. Risks in projects are observable across various types of projects. In civil engineering and construction projects, risks can impact time, productivity, quality, and the necessary project budget. Similarly, in research and development projects, one can witness risks such as incomplete work, product failure, market unpredictability, and more (Ceyhun, 2017). The most significant effects of these risks include failure to complete the project within the allocated budget, timeframe, and achieving the required quality. The objective of risk management operations is to mitigate the involved parties from these impacts (Schnetler et al., 2015). Considering the widespread influence that different risks can have on project execution, since the 1990s, numerous authors have proposed various processes for risk management in this domain (Rahman & Adnan, 2020). In all these proposed processes, three main activities exist: identification, analysis, and response to risks. Identifying the risks affecting projects is crucial and will be the first significant step (De Marco & Thaheem, 2014). This identification can be regarded as one of the most vital stages in risk management. Project risk management encompasses processes that involve guiding risk management planning, identification, analysis, response, and monitoring and control throughout the project. Often, these processes are updated during the project lifecycle. The objectives of project risk management include reducing the likelihood of adverse events affecting the project (Ferreira de Araújo Lima et al., 2021).

Life continues amidst conditions of uncertainty that cast shadows over all matters and transform the decision-making process. Nowadays, virtually all operational activities and processes are viewed through the lens of risk (Fourie, 2022). The ultimate strategic plan must create the capacity to embrace more risks, as it is the only way to enhance performance (Jayasudha & Vidivelli, 2016). To expand this capacity, companies must recognize the risks they are willing to accept. Instead of immersing themselves in uncertainty based on guesswork or hearsay, companies must be able to rationally choose their risks. Risk management is a crucial Endeavor (Tyagi, 2020). If initiated timely within a project's activities, it can prove beneficial and serve as a powerful tool for early identification of weaknesses (Kishore, Pretorius, & Chattopadhyay, 2019). This allows the management team to organize operational plans to manage risks, prevent them from becoming significant issues in the future, and, thus, your proactive response to potential issues can save both money and time (Naik & Prasad, 2021).

## 2. Methodology

The research method based on its objective can be fundamental and practical. Considering that the desired results of this study can be applied in the selection of project managers, our research is applied in nature. Applied Research: It is a type of research that uses the results of fundamental research to improve and enhance behaviours, methods, tools, materials, structures, and patterns used in human societies. The goal of applied research is to develop practical knowledge in a specific field. In this research, a library research method was employed, involving online searches and consulting articles and relevant books in the field of managerial and informational risks. Past research findings were explored and potential risks affecting civil engineering project management were identified. After formulating the questionnaire, key indicators and factors were determined under the supervision of relevant experts. This research, based on the method of obtaining necessary data, is descriptive (non-experimental) and of a survey type. Descriptive research includes a set of methods whose goal is to describe conditions or phenomena under investigation. Conducting descriptive research can be solely for understanding existing conditions or assisting in the decision-making process. Most behavioural science research falls under the category of descriptive research. Descriptive research is further divided into various types, including survey, correlation, action research, case study, and post-event research, which examines the distribution and characteristics of the population and investigates the nature of existing conditions and the relationship between events.

Population

Identifying the population in research is of great importance because using the population, the sample size is determined, and then statistical analyses are conducted based on the sample. The population of this research consists of experts in the field of civil engineering project management. In this study, information required for the research was obtained through interaction with 7 experts from this population, with coordination in place.

Data Collection Method

There are multiple methods available for data collection, and often, in a research study, more than one method is utilized to gather information. In this research, data was collected through field methods. In field methods, the questionnaire is one of the most common tools for data collection. In preparing the questionnaire, the researcher attempts to obtain necessary information from the respondents through a series of questions.

Data Collection Tools

Questionnaire is one of the most common tools for collecting information in survey research. In this study, a questionnaire was used as the data collection tool. The questionnaire's question numbers related to the FMEA method, along with the corresponding questions, are provided in the table below.

Considered Risks Along with Corresponding Questions

| Risks | Related Questions |
|---|---|
| Individual Skills Risk | 1 to 2 |
| Power-Related Risks | 3 to 7 |
| Knowledge and Expertise Risks | 8 to 10 |
| Experience Risks | 11 to 13 |
| Personality Traits Risks | 14 to 17 |

After filling in the relevant cells (severity, occurrence, and detection) by the experts, the information is used to calculate the modified RPN values, and the most significant risks are identified. The severity, occurrence, and detection rating tables are provided below.

Severity Rating Scale

| Rank | Severity Level | Description |
|---|---|---|
| 10 | Hazardous – No Warning | Extremely severe, catastrophic |
| 9 | Hazardous – Warning | Extremely severe but with a warning |
| 8 | Very High | Irrecoverable severity |
| 7 | High | High severity |
| 6 | Moderate | Moderate severity |
| 5 | Low | Low severity |
| 4 | Very Low | Very low severity but most individuals notice it |
| 3 | Minor Effects | Causes minor effects |
| 2 | Very Minor | Very minor effects |
| 1 | None | No effect |

Occurrence Rating Scale

| Likelihood of Risk Occurrence | Probability Rates of Risk | Rank |
|---|---|---|
| Very High – Almost Inevitable | 1 in 2 or more | 10 |
| 1 in 3 | 9 | |
| High: Repeating risks | 1 in 8 | 8 |
| 1 in 20 | 7 | |
| Moderate: Occasional risks | 1 in 80 | 6 |
| 1 in 400 | 5 | |
| 1 in 2000 | 4 | |
| Low: Relatively rare risks | 1 in 15000 | 3 |
| 1 in 1500000 | 2 | |
| Remote: Highly improbable risk | Less than 1 in 15000000 | 1 |

Detection Rating Scale

| Criteria: Risk Detection Capability | Detection Ability | Rank |
|---|---|---|
| No controls exist, or if they do, they cannot detect the potential risk | Absolutely None | 10 |
| Very unlikely that the risk will be detected even with existing controls | Very Unlikely | 9 |
| Highly unlikely that the risk will be detected even with existing controls | Unlikely | 8 |
| Very low likelihood that the risk will be detected even with existing controls | Very Low | 7 |
| Low likelihood that the risk will be detected even with existing controls | Low | 6 |
| Possible in half of the cases that the potential risk will be detected with existing controls | Moderate | 5 |
| Relatively likely that the potential risk will be detected with existing controls | Relatively Likely | 4 |
| Very likely that the potential risk will be detected with existing controls | Likely | 3 |
| Highly likely that the potential risk will be detected | Very Likely | 2 |
| Almost certain that the potential risk will be detected with existing controls | Almost Certain | 1 |

After formulating the initial questionnaire, efforts are made to determine the validity and reliability of the questionnaire.

3. Results

The reliability of the questionnaire used in this study was examined by calculating Cronbach's alpha. If Cronbach's alpha is greater than 0.7, the questionnaire is considered reliable.

Cronbach's Alpha: 0.702, Number of Items: 17

As observed, the Cronbach's alpha of the questionnaire used in this study is 0.702, indicating the research tool's reliability.

Model Construction

The model presented in this study is explained at three levels. The objective is to determine the prioritization weight of risk-generating parameters: severity, occurrence, and detection. The second level is the cluster level, termed the "Failure Clusters." Since each potential failure scenario forms a failure cluster, here we have five failure clusters named individual skills risk, power risks, knowledge and expertise risks, experience risks, and personality traits risks. These clusters along with their elements are described in the table below.

Failure Clusters of the Model and Their Elements

| Failure Cluster | Elements |
| --- | --- |
| Individual Skills Risk | Inability to justify subordinates |
|  | Lack of decisive speech |
| Power Risks | Insufficient power in the position |
|  | Lack of power of exploitation |
|  | Lack of expert power |
|  | Lack of information power |
|  | Lack of political power |
| Knowledge and Expertise Risks | Lack of management knowledge |
|  | Lack of civil engineering knowledge |
|  | Lack of executive knowledge |
| Experience Risks | Lack of construction experience |
|  | Lack of management experience |
|  | Lack of project management experience |
| Personality Traits Risks | Lack of humility |
|  | Lack of good ethics |
|  | Lack of trustworthiness |
|  | Lack of perseverance |

This level is the strategic level; it plays a fundamental role in decision-making and serves as an intermediary between the first and third levels. At the third level, the model's options are described, consisting of severity, occurrence, and detection. The next step involves examining the types of dependencies and comparisons in the sampled data. Identification of dependencies and pairwise comparisons was carried out by the group members' consensus.

Determining Dependencies and Pairwise Comparison Matrices

In this stage, comparative matrices of main criteria, the interdependence of main criteria, sub-criteria, and their interdependencies are formed, and their consistency is controlled. The following steps are explained:

Binary Comparison of Main Criteria with Respect to the Objective

The binary comparison of the five main criteria is performed based on the 9-point quantitative hourly scale, in the same order used in the Analytic Hierarchy Process (AHP) hierarchical analysis process. The result of the binary comparison of the main criteria is presented in the table below.

Binary Comparison of Main Criteria with Respect to the Objective

|  | Individual Skills | Power | Knowledge & Expertise | Experience | Personality Traits |
|---|---|---|---|---|---|
| Individual Skills | 1 | 1/3 | 1/7 | 1/9 | 1/2 |
| Power |  | 1 | 1/3 | 1/5 | 5 |
| Knowledge & Expertise |  |  | 1 | 1/2 | 5 |
| Experience |  |  |  | 1 | 7 |
| Personality Traits |  |  |  |  | 1 |

Inconsistency Ratio: 0.01097

Binary Comparison of Internal Dependencies of Main Criteria

To understand the mutual dependencies between the main criteria, binary comparisons are made between the main criteria based on the 9-point hourly quantitative scale. The method for calculating the importance coefficient of each of the main criteria (considering their mutual interdependence) is illustrated through pairwise comparisons of the five main criteria while controlling each of the criteria, as presented in the tables below. The questioning method for the importance coefficient in these cases is as follows: How important is criterion A relative to B when C is controlled?

Internal Dependency Matrix of Main Criteria

|  | Individual Skills | Power | Knowledge & Expertise | Experience | Personality Traits |
|---|---|---|---|---|---|
| Individual Skills | ✓ | ✓ | ✓ | ✓ | ✓ |
| Power | ✓ | ✓ | ✓ | ✓ | ✓ |
| Knowledge & Expertise | ✓ | ✓ | ✓ | ✓ | ✓ |
| Experience | ✓ | ✓ | ✓ | ✓ | ✓ |
| Personality Traits | ✓ | ✓ | ✓ | ✓ | ✓ |

Binary comparison of the main criteria considering their internal dependencies with controlling Individual Skills

|  | Power | Knowledge & Expertise | Experience | Personality Traits |
|---|---|---|---|---|
| Power | 1 | 1/3 | 1/3 | 3 |
| Knowledge & Expertise |  | 1 | 1/2 | 5 |
| Experience |  |  | 1 | 7 |
| Personality Traits |  |  |  | 1 |

Inconsistency Ratio: 0.03901

Binary comparison of the main criteria considering their internal dependencies with controlling Power

|  | Individual Skills | Knowledge & Expertise | Experience | Personality Traits |
|---|---|---|---|---|
| Individual Skills | 1 | 1/3 | 1/5 | 1 |
| Knowledge & Expertise |  | 1 | 1/3 | 3 |
| Experience |  |  | 1 | 5 |
| Personality Traits |  |  |  | 1 |

Inconsistency Ratio: 0.01629

Binary comparison of the main criteria considering their internal dependencies with controlling Knowledge & Expertise

|  | Individual Skills | Power | Experience | Personality Traits |
|---|---|---|---|---|
| Individual Skills | 1 | 1/5 | 1/9 | 1/2 |
| Power |  | 1 | 1/5 | 3 |
| Experience |  |  | 1 | 7 |
| Personality Traits |  |  |  | 1 |

Inconsistency Ratio: 0.04408

Binary comparison of the main criteria considering their internal dependencies with controlling Experience

|  | Individual Skills | Power | Knowledge & Expertise | Personality Traits |
|---|---|---|---|---|
| Individual Skills | 1 | 2/1 | 7/1 | 3/1 |
| Power |  | 1 | 5/1 | 2/1 |
| Knowledge & Expertise |  |  | 1 | 3 |
| Personality Traits |  |  |  | 1 |

Inconsistency Ratio: 0.00719

Binary Comparison of Sub-Criteria of Each Main Criterion

At this stage, the importance coefficient of each of the sub-criteria related to the five main criteria is obtained through their pairwise binary comparisons (based on the 9-point hourly quantitative scale). The results of the pairwise binary comparisons of the sub-criteria related to each of the main criteria are presented in the tables below.

Binary comparison of sub-criteria related to Individual Skills

|  | Inability to Justify Subordinates | Lack of Assertiveness |
|---|---|---|
| Inability to Justify Subordinates | 1 | 3 |
| Lack of Assertiveness |  | 1 |

Inconsistency Ratio: N/A

Binary comparison of sub-criteria related to Power

|  | Lack of Adequate Authority in Position | Lack of Operational Authority | Lack of Expert Authority | Lack of Informational Authority | Lack of Political Authority |
|---|---|---|---|---|---|
| Lack of Adequate Authority in Position | 1 | 3 | 3/1 | 2 | 2 |
| Lack of Operational Authority |  | 1 | 7/1 | 3/1 | 3/1 |
| Lack of Expert Authority |  |  | 1 | 5 | 5 |
| Lack of Informational Authority |  |  |  | 1 | 1 |
| Lack of Political Authority |  |  |  |  | 1 |

Inconsistency Ratio: 0.02032

Binary Comparison of Internal Dependencies of Sub-Criteria

As stated in Chapter 3, 17 sub-criteria (potential failure causes) representing the characteristics of the five main criteria have been selected for the purposes of this study. The mutual dependencies of these sub-criteria are shown in the table below. To arrive at this table and determine the mutual dependencies of the sub-criteria, the opinions of relevant experts have been utilized.

Internal Dependencies of Sub-Criteria

| Sub-Criteria | Lack of Sufficient Power in Position | Lack of Management Knowledge | Lack of Managerial Experience | Lack of Good Ethics | Lack of Perseverance | Lack of Execution Knowledge | Lack of Project Management Experience | Lack of Civil Engineering Knowledge | Lack of Expertise |
|---|---|---|---|---|---|---|---|---|---|
| Inability to Justify Subordinates | ✓ | ✓ | ✓ | | | | | | |
| Lack of Definitive Speech | ✓ | | ✓ | | | | | | |
| Lack of Sufficient Power in Position | | ✓ | | ✓ | ✓ | | | | |
| Lack of Operational Authority | | | | | | ✓ | ✓ | | |
| Lack of Expertise | | | | | | | | ✓ | |
| Lack of Managerial Experience | | | | | ✓ | | ✓ | ✓ | ✓ |
| Lack of Project Management Experience | ✓ | | ✓ | ✓ | | ✓ | | | ✓ |
| Lack of Civil Engineering Knowledge | | | ✓ | | | | | | |
| Lack of Expert Authority | | | | | | | ✓ | | ✓ |

Binary comparison of sub-criteria with mutual dependencies with the sub-criterion Inability to Justify Subordinates

| Sub-Criteria | Lack of Sufficient Power in Position | Lack of Management Knowledge | Lack of Managerial Experience |
|---|---|---|---|
| Lack of Sufficient Power in Position | 1 | 3/1 | 2 |
| Lack of Management Knowledge | | 1 | 5 |
| Lack of Managerial Experience | | | 1 |

Inconsistency Ratio: 0.00355

Binary comparison of sub-criteria with mutual dependencies with the sub-criterion Lack of Definitive Speech

| Sub-Criteria | Lack of Sufficient Power in Position | Lack of Managerial Experience |
|---|---|---|
| Lack of Sufficient Power in Position | 1 | 3 |
| Lack of Managerial Experience | | 1 |

Inconsistency Ratio: 0.00

Binary comparison of sub-criteria with mutual dependencies with the sub-criterion Lack of Sufficient Power in Position

| Sub-Criteria | Lack of Management Knowledge | Lack of Good Ethics | Lack of Perseverance |
|---|---|---|---|
| Lack of Management Knowledge | 1 | 2/1 | 2 |

| | | 1 | 5 |
|---|---|---|---|
| Lack of Good Ethics | | 1 | 5 |
| Lack of Perseverance | | | 1 |

Inconsistency Ratio: 0.00532

Binary comparison of sub-criteria with mutual dependencies with the sub-criterion Lack of Operational Authority

| Sub-Criteria | Lack of Execution Knowledge | Lack of Project Management Experience |
|---|---|---|
| Lack of Execution Knowledge | 1 | 3/1 |
| Lack of Project Management Experience | | 1 |

Inconsistency Ratio: 0.00

Binary comparison of sub-criteria with mutual dependencies with the sub-criterion Lack of Expert Authority

| Sub-Criteria | Lack of Management Knowledge | Lack of Civil Engineering Knowledge |
|---|---|---|
| Lack of Management Knowledge | 1 | 2/1 |
| Lack of Civil Engineering Knowledge | | 1 |

Inconsistency Ratio: 0.00

Binary comparison of sub-criteria with mutual dependencies with the sub-criterion Lack of Project Management Experience

| Sub-Criteria | Lack of Management Knowledge | Lack of Execution Knowledge | Lack of Expert Authority |
|---|---|---|---|
| Lack of Management Knowledge | 1 | 7 | 5 |
| Lack of Execution Knowledge | | 1 | 3/1 |
| Lack of Expert Authority | | | 1 |

Inconsistency Ratio: 0.06239

Binary Comparison of Option Preferences

At this stage, the preference of each option in relation to each sub-criterion is examined and judged. The basis for this judgment is the same 9-point Likert scale, with the difference that the comparison of options in relation to each sub-criterion discusses the preference of options rather than their importance.

Binary comparison of option preferences in relation to the sub-criterion Inability to Justify Subordinates

| Intensity | Occurrence | Identification |
|---|---|---|
| 1 | 3 | 5 |
| | 1 | 2 |
| | | 1 |

Inconsistency Ratio: 0.00355

Binary comparison of option preferences in relation to the sub-criterion Lack of Definitive Speech

| Intensity | Occurrence | Identification |
|---|---|---|
| 1 | 5/1 | 3 |
| | 1 | 7/1 |
| | | 1 |

Inconsistency Ratio: 0.06239

Binary comparison of option preferences in relation to the sub-criterion Lack of Sufficient Power in Position

| Intensity | Occurrence | Identification |
|---|---|---|
| 1 | 3 | 3/1 |

|  | 1 | 5/1 |
|---|---|---|
|  |  | 1 |

Inconsistency Ratio: 0.03703

Binary comparison of option preferences in relation to the sub-criterion Lack of Operational Authority

| Intensity | Occurrence | Identification |
|---|---|---|
| 1 | 3/1 | 5/1 |
|  | 1 | 2/1 |
|  |  | 1 |

Inconsistency Ratio: 0.00355

Binary comparison of option preferences in relation to the sub-criterion Lack of Expert Authority

| Intensity | Occurrence | Identification |
|---|---|---|
| 1 | 2 | 4 |
|  | 1 | 3 |
|  |  | 1 |

Inconsistency Ratio: 0.01759

Binary comparison of option preferences in relation to the sub-criterion Lack of Information Power

| Intensity | Occurrence | Identification |
|---|---|---|
| 1 | 1 | 3/1 |
|  | 1 | 3/1 |
|  |  | 1 |

Inconsistency Ratio: 0.00

Binary comparison of option preferences in relation to the sub-criterion Lack of Political Power

| Intensity | Occurrence | Identification |
|---|---|---|
| 1 | 5 | 2 |
|  | 1/2 | 1 |
|  |  | 1 |

Inconsistency Ratio: 0.00532

Binary comparison of option preferences in relation to the sub-criterion Lack of Management Knowledge

| Intensity | Occurrence | Identification |
|---|---|---|
| 1 | 2 | 4 |
|  | 1 | 2 |
|  |  | 1 |

Inconsistency Ratio: 0.00

Binary comparison of option preferences in relation to the sub-criterion Lack of Civil Engineering Knowledge

| Intensity | Occurrence | Identification |
|---|---|---|
| 1 | 5 | 7 |
|  | 1 | 3 |
|  |  | 1 |

Inconsistency Ratio: 0.06239

Binary comparison of option preferences in relation to the sub-criterion Lack of Execution Knowledge

| Intensity | Occurrence | Identification |
|---|---|---|
| 1 | 1/2 | 3 |
|  | 1 | 5 |
|  |  | 1 |

Inconsistency Ratio: 0.00355

Binary comparison of option preferences in relation to the sub-criterion Lack of Construction Experience

| Intensity | Occurrence | Identification |
|---|---|---|
| 1 | 5 | 5 |
|   | 1 | 1 |
|   |   | 1 |

Inconsistency Ratio: 0.00

Binary comparison of option preferences in relation to the sub-criterion Lack of Managerial Experience

| Intensity | Occurrence | Identification |
|---|---|---|
| 1 | 2 | 5 |
|   | 1 | 2 |
|   |   | 1 |

Inconsistency Ratio: 0.00532

Binary comparison of option preferences in relation to the sub-criterion Lack of Project Management Experience

| Intensity | Occurrence | Identification |
|---|---|---|
| 1 | 3 | 3 |
|   | 1 | 1 |
|   |   | 1 |

Inconsistency Ratio: 0.00

Binary comparison of option preferences in relation to the sub-criterion Lack of Humility

| Intensity | Occurrence | Identification |
|---|---|---|
| 1 | 1/3 | 1/5 |
|   | 1 | 3 |
|   |   | 1 |

Inconsistency Ratio: 0.03703

Binary comparison of option preferences in relation to the sub-criterion Lack of Good Ethics

| Intensity | Occurrence | Identification |
|---|---|---|
| 1 | 2 | 1/5 |
|   | 1 | 1/7 |
|   |   | 1 |

Inconsistency Ratio: 0.01361

Binary comparison of option preferences in relation to the sub-criterion Lack of Trustworthiness

| Intensity | Occurrence | Identification |
|---|---|---|
| 1 | 3 | 1/3 |
|   | 1 | 1/5 |
|   |   | 1 |

Inconsistency Ratio: 0.03703

Binary comparison of option preferences in relation to the sub-criterion Lack of Perseverance

| Intensity | Occurrence | Identification |
|---|---|---|
| 1 | 1 | 1/5 |
|   | 1 | 1/5 |
|   |   | 1 |

Inconsistency Ratio: 0.00

Formation of Super matrix and Calculation of Parameter Weights

After pairwise comparisons among decision statements (clusters and elements) and placing the relative weights of these statements in a unit matrix, an initial non-normalized super matrix is formed. The rows and columns of this matrix represent clusters and their corresponding elements. The components of this matrix indicate the weight of the corresponding element in the row to the element in the column. After forming the non-normalized super matrix, it needs to be normalized. In the normalized super matrix, the sum of each

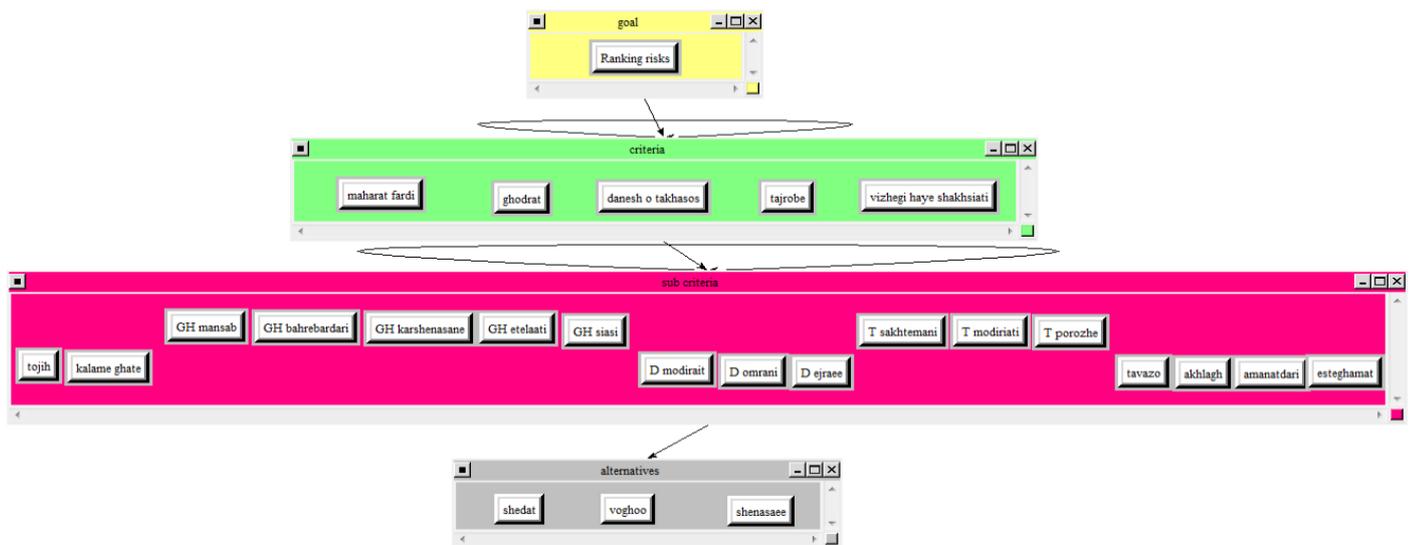

column will be equal to one. To achieve this, the elements in each block of the non-normalized super matrix are multiplied by the weight of that block. In the final step, the obtained normalized super matrix is elevated to high powers until no significant difference is observed between the components of its rows. This final super matrix is referred to as the ultimate super matrix. In this study, the steps related to the ANP method were performed using the SUPER DECISION software. The figures below show the results of the software and also the model diagram written in the software. The results obtained from the comparisons in the software. Values 0.55, 0.22, and 0.23, respectively, correspond to the intensity, identification, and occurrence rows of option clusters, the unnormalized weights of the parameters. Using the correction formula, the parameter weights will be as follows: Occurrence = 0.69 Identification = 0.66 Intensity = 1.65

Problem Structuring Method (PSM) in Super Decision Software

Calculation of RPN for Each Identified Factor

To calculate the risk priority number (RPN) for each of the identified factors, the priority score for the severity, occurrence, and detection parameters, considering the weights obtained for each, has been calculated using the new RPN formula.

Calculation of Risk Priority and Determination of Priority for Identified Failure Factors

| Row | Potential Failure Mode | Potential Failure Causes | RPN | Priority |
|---|---|---|---|---|
| 1 | Risk of Individual Skills | Inability to Justify Subordinates | 267.9974 | 8 |
| | | Lack of Decisive Speech | 306.1735 | 7 |
| 2 | Power Risks | Insufficient Positional Power | 254.6083 | 10 |
| | | Lack of Operational Power | 161.8835 | 15 |
| | | Lack of Expert Power | 307.8528 | 6 |
| | | Lack of Informational Power | 322.689 | 5 |
| | | Lack of Political Power | 555.348 | 1 |

| 3 | Knowledge and Expertise Risks | Lack of Management Knowledge | 123.405 | 16 |
| | | Lack of Civil Engineering Knowledge | 89.22567 | 17 |
| | | Lack of Executive Knowledge | 265.6944 | 9 |
| 4 | Experience Risks | Lack of Construction Experience | 514.2947 | 2 |
| | | Lack of Management Experience | 201.5771 | 12 |
| | | Lack of Project Management Experience | 466.934 | 3 |
| 5 | Personality Traits Risks | Lack of Humility | 191.261 | 13 |
| | | Lack of Good Ethics | 253.3361 | 11 |
| | | Lack of Trustworthiness | 176.9562 | 14 |
| | | Lack of Perseverance | 415.8512 | 4 |

For example, in the first case, the risk priority number for the inability to justify subordinates is calculated using the following formula:

$$267.9974 = 8^{1.65} \times 8^{0.66} \times 3^{0.69}$$

Corrective Actions Implementation

Corrective actions involve activities aimed at eliminating or reducing potential failure factors. Clearly, in making decisions about corrective actions, priority is given to factors with higher risk priority. After addressing these, it will be logical to deal with lower-priority risk factors. According to the results obtained in this study, the risk factors of lack of political power, lack of construction experience, and lack of project management experience have been identified as the most critical risks in selecting a project manager.

4. **Discussion**

In this research, we present an approach to risk analysis for selecting project managers in civil engineering projects, utilizing a combination of Failure Modes and Effects Analysis (FMEA) and the Analytic Network Process (ANP) methodology. Initially, we conducted a comprehensive literature review to understand existing research in the field. Following this, we identified the risks associated with selecting project managers in civil engineering projects and analysed these identified risks using the integrated ANP-FMEA approach. This section provides an overview of the overall work conducted, followed by an analysis of the results using a practical approach. The content presented in this section encapsulates the findings of the conducted research. The ANP-FMEA method is an advanced risk analysis approach with a simple and ingenious logic. This method is built on the foundation of the FMEA technique and has been modified to be more compatible with the real world. By combining these methods, the ANP-FMEA approach adds significant value to the risk management process. Failure to consider the interrelationships between risk factors and assigning equal weights to FMEA parameters can significantly reduce the effectiveness of corrective actions. This drawback is addressed by the ANP method. ANP-FMEA extends the simple concept of risk priority numbers (RPN) and assigns different importance values to FMEA parameters using powers. The calculated RPN values obtained under the system conditions in which it is employed show better compatibility.

The table below illustrates the RPN values and priority of potential failure factors in both FMEA and ANP-FMEA methods, highlighting the differences and alignments between the two approaches

| Potential Failure Factors | RPN (FMEA) | RPN (ANP-FMEA) | Priority (FMEA) | Priority (ANP-FMEA) |
|---|---|---|---|---|
| Inability to justify subordinates | 192 | 268 | 12 | 8 |
| Lack of decisive speech | 240 | 306 | 8 | 7 |
| Insufficient power in position | 224 | 255 | 10 | 10 |
| Lack of operational power | 144 | 162 | 15 | 15 |
| Lack of expertise power | 240 | 308 | 8 | 6 |
| Lack of informational power | 216 | 323 | 11 | 5 |
| Lack of political power | 576 | 555 | 1 | 1 |
| Lack of managerial knowledge | 120 | 123 | 16 | 16 |
| Lack of civil engineering knowledge | 75 | 89.23 | 17 | 17 |
| Lack of executive knowledge | 192 | 266 | 12 | 9 |
| Lack of construction experience | 432 | 514 | 3 | 2 |

| Lack of managerial experience | 192 | 201.58 | 12 | 12 |
| Lack of project management experience | 448 | 466.93 | 2 | 3 |
| Lack of humility | 324 | 191.26 | 4 | 13 |
| Lack of good ethics | 270 | 253.34 | 7 | 11 |
| Lack of trustworthiness | 288 | 176.96 | 6 | 14 |
| Lack of perseverance | 315 | 415.85 | 5 | 4 |

Upon observing the table above, the following findings can be articulated regarding the differences and alignments between FMEA and ANP-FMEA methods: Most RPN values in ANP-FMEA (except for lack of humility, lack of good ethics, and lack of trustworthiness) are larger than those in FMEA. This indicates that when proper weights are applied to each parameter, their risk levels are higher than perceived. For higher priorities (1, 2, etc.), the results of both methods are nearly identical. This means that ANP-FMEA tends to have lower risk acceptance for lower priorities, or in other words, it seeks factors whose mitigation provides greater assurance. The sensitivity of ANP-FMEA in distinguishing potential failure factors' priorities is higher. In FMEA, three factors share priority 12 (inability to justify subordinates, lack of executive knowledge, and lack of managerial experience) and two share priority 8 (lack of decisive speech, lack of operational power). In contrast, in ANP-FMEA, no two factors share the same rank. ANP-FMEA can thus differentiate failure factors more effectively than FMEA. Additionally, ANP-FMEA can create classifications for RPN based on the industry-specific weights employed. This means that depending on the specific industry, separate categories for RPN can be considered for more tailored and effective corrective actions.

Limitations of the Study

FMEA is a group-oriented method that requires collaboration, awareness, and engagement from a group of stakeholders, which can be a limitation in terms of coordination and may create constraints in the method's implementation. Another fundamental limitation is the reliance on the senior management and project owners' participation and commitment. The success of this process significantly depends on their decision-making for the identification and reduction of risks associated with selecting a project manager.

In this research, an attempt has been made to address the limitations of the common FMEA method by employing the ANP method. As discussed earlier, several limitations of the conventional FMEA method were outlined by stakeholders. Given the extensive use of the FMEA method, any effort towards its evolution can be invaluable. Additionally, in some cases, experts may not have a clear understanding of the importance levels. To resolve this issue, it is suggested to use a solution of fuzzy trapezoidal numbers. This solution can allow experts to express their opinions in various ways: precise numbers, a range of numbers, linguistic terms, fuzzy numbers, or no opinion. By using this method, disparate systems of comparison can ultimately converge into a unified system.

Future Research

The presented priorities of risk factors can serve as a numerical framework for selecting project managers. These coefficients can be a useful guide for senior managers of civil engineering companies when lacking sufficient experience in selecting a suitable project manager. However, it should be noted that these coefficients are based on the opinions of managers of Iranian companies. Naturally, managers' opinions in other countries and different conditions might vary. Furthermore, these coefficients are calculated for civil engineering projects. Calculating the importance coefficients of project manager selection criteria for other types of projects, such as oil and gas projects, could be a subject for future research. Different projects require project managers with different abilities and skills. To elucidate the applicability and effectiveness of the proposed method, it is suggested to implement the model as a practical case study for selecting a project manager in a civil engineering project. Such a study can demonstrate the efficiency and effectiveness of the proposed method in practical scenarios. The method presented in this research serves as a valuable tool for decision-making for organizational senior managers. Apart from its specific application in selecting project managers, it has the potential for generalization and implementation in other management decision-making issues, such as analysing the risk of selecting suppliers and contractors in supply chain management. For future research, the utilization of fuzzy operators for calculations and incorporating parameters other than probability, identification, and severity in risk calculations is recommended.

## 5. Conclusion

In the realm of civil engineering projects, selecting the right project manager is a critical task. Traditional methods involve candidate assessments, but the criteria for this selection are complex, varying based on project type and associated risks. This research delved deep into this complexity, identifying five key risk categories: individual skills, power-related issues, knowledge and expertise, experience, and personality traits. Employing the combined ANP-FMEA approach, our study pinpointed significant risks in the selection process. Notably, the lack of political influence, absence of construction experience, and deficiency in project management expertise emerged as crucial pitfalls. What sets our research apart is the innovative use of ANP-FMEA, a method that excels in differentiating risks. It offers nuanced insights, aiding senior managers in making informed decisions. In essence, this study not only illuminates the risks in selecting project managers but also provides a robust tool for navigating these challenges. The ANP-FMEA model, with its detailed risk differentiation, empowers senior managers, ensuring the choice of project managers tailored to specific project needs. As the construction industry advances, these insights and methodologies stand as invaluable assets, ensuring project success and organizational growth.

**References**

Agbejule, A., & Lehtineva, L. (2022). The relationship between traditional project management, agile project management and teamwork quality on project success. International Journal of Organizational Analysis, 30(7), 124-136.

Ahmed, S., & El-Sayegh, S. (2020). Critical review of the evolution of project delivery methods in the construction industry. Buildings, 11(1), 11.

Ahmadi-Javid, A., Fateminia, S. H., & Gemünden, H. G. (2020). A method for risk response planning in project portfolio management. Project Management Journal, 51(1), 77-95.

Cakmak, P. I., & Tezel, E. (2019). A guide for risk management in construction projects: Present knowledge and future directions. Risk management in construction projects.

Ceyhun, G. Ç. (2017). Risk Management Practices in Strategic Management. Global Business Strategies in Crisis: Strategic Thinking and Development, 263-271.

Daneshmandi, F., Hessari, H., Nategh, T., & Bai, A. (2023). Examining the Influence of Job Satisfaction on Individual Innovation and Its Components: Considering the Moderating Role of Technostress. arXiv preprint arXiv:2310.13861.

De Marco, A., & Thaheem, J. (2014). Risk analysis in construction projects: a practical selection methodology.

Ferreira de Araújo Lima, P., Marcelino-Sadaba, S., & Verbano, C. (2021). Successful implementation of project risk management in small and medium enterprises: a cross-case analysis. International Journal of Managing Projects in Business, 14(4), 1023-1045.

Fourie, W. (2022). Leadership and risk: a review of the literature. Leadership & Organization Development Journal, 43(4), 550-562.

Hessari, H., Busch, P., & Smith, S. (2022a). Supportive leadership and co-worker support for nomophobia reduction: Considering affective commitment and HRM practices. ACIS 2022 Proceedings. 18. https://aisel.aisnet.org/acis2022/18

Hessari, H., & Nategh, T. (2022b). The role of co-worker support for tackling techno stress along with these influences on need for recovery and work motivation. International Journal of Intellectual Property Management, 12(2), 233-259. https://doi.org/10.1504/IJIPM.2022.122301

Hessari, H., & Nategh, T. (2022c). Smartphone addiction can maximize or minimize job performance? Assessing the role of life invasion and techno exhaustion. Asian Journal of Business Ethics, 11(1), 159-182. https://doi.org/10.1007/s13520-022-00145-2

Jayasudha, K., & Vidivelli, B. (2016). Analysis of major risks in construction projects. ARPN journal of engineering and applied sciences, 11(11), 6943-6950.

Kishore, N., Pretorius, J. H. C., & Chattopadhyay, G. (2019, December). The Roles of Functional Managers and Project Managers in a Matrix Organization. In 2019 IEEE International Conference on Industrial Engineering and Engineering Management (IEEM) (pp. 784-788). IEEE.

Naik, S., & Prasad, C. V. (2021). Benefits of enterprise risk management: A systematic review of literature. Reference to this paper should be made as follows: Naik, S, 28-35.

Rahman, M., & Adnan, T. (2020). Risk management and risk management performance measurement in the construction projects of Finland. Journal of Project Management, 5(3), 167-178.

Schnetler, R., Steyn, H., & Van Staden, P. J. (2015). Characteristics of matrix structures, and their effects on project success. South African Journal of Industrial Engineering, 26(1), 11-26.

Tyagi, A. (2020). Enterprise Risk Management: Benefits and Challenges. Available at SSRN 3748267.

Zhu, F., Hu, H., & Xu, F. (2022). Risk assessment model for international construction projects considering risk interdependence using the DEMATEL method. Plos one, 17(5), e0265972.